\documentclass[11pt]{article}
\usepackage[centertags]{amsmath}
\usepackage{amsfonts}
\usepackage{newlfont}
\usepackage{graphicx}
\usepackage{eurosym}
\usepackage[dvips]{color}
\usepackage{epstopdf}

\newtheorem{lemma}{Lemma}[section]
\newtheorem{remark}{Remark}[section]
\newcommand{\F}{\cal{F}}

\newcommand{\G}{\cal{G}}

\newcommand{\Fu}{{\cal{F}}_u}

\newcommand{\Ft}{{\cal{F}}_t}
\newcommand{\Gs}{{\cal{G}}_s}

\newcommand{\Gt}{{\cal{G}}_t}
\newcommand{\Ht}{{\cal{H}}_t}

\newcommand{\E}{\mathbf{E}}

\newcommand{\e}{\mathrm{e}}

\newcommand{\lam}{\lambda}
\newcommand{\cc}{\mathbf{c}}

\newcommand{\ds}{\displaystyle}

\newcommand{\cA}{\cal A}
\newcommand{\cF}{\cal F}
\newcommand{\cL}{\cal L}

\begin{document}
\title{CVA and vulnerable options pricing  by correlation expansions}

\author{F. Antonelli\footnote{ University of L'Aquila,
          \texttt{fabio.antonelli@univaq.it}}, A. Ramponi\footnote{Dept. Economics and Finance, University of Roma - Tor Vergata, \texttt{alessandro.ramponi@uniroma2.it}}, S. Scarlatti\footnote{Dept. Enterprise Engineering, University of Roma - Tor Vergata, \texttt{sergio.scarlatti@uniroma2.it}}}



\maketitle


\begin{abstract}
We consider the problem of computing the Credit Value Adjustment ({CVA}) of a European option in presence of the Wrong Way Risk ({WWR}) in a default intensity setting. Namely we model the asset price evolution as solution to a linear equation that might depend on different stochastic factors and we provide an approximate evaluation of the option's price, by exploiting a correlation expansion approach, introduced in \cite{AS}. We compare the numerical performance of such a method with that recently proposed by Brigo et al. (\cite{BR18}, \cite{BRH18}) in the case of a call option driven by a GBM correlated with the CIR default intensity. We additionally report some  numerical evaluations obtained by other methods.

\medskip



\end{abstract}

\section{Introduction}

Vulnerable options are financial contracts that are subject to some default event concerning the solvability of the option's  seller. The classical reference on this topic is  the paper by Johnson and Stulz, \cite{JS87},  the first to price European options with Counterparty Credit Risk ({CCR}). Their work was developed within the structural approach to credit risk and it considered  the option as the sole liability of the counterparty. Later Klein,  in \cite{Kl96},  discussed  more general liability structures and the presence of correlation between the option's underlying and the option's seller's assets, while in \cite{KIn99} interest rate risk was included and in   \cite{KIn01}  a (stochastic) default barrier depending on the value of the option was considered. In all these works default could happen only at maturity.

In the meantime reduced-form models to price bonds or options that might default at any time prior to maturity, started to be proposed.  We refer the reader to Hull and White (\cite{HW95})  and Jarrow and Turnbull (\cite{JT95}) for the case of vulnerable options and to   \cite{DS99}  and the references therein for a more  general framework. Some more recent papers on vulnerable options include \cite{CL03}, \cite{CPV14}, \cite{TW14}, \cite{Fard15} and \cite{Kao16}.

Even before the last financial crisis (2007-2008), the focus on {CCR} started to increase notably (see \cite{CD03}) and attention shifted to building a general framework for the evaluation of a premium to compensate  a derivative's holder  (in particular of Interest Rate Swaps) for taking (counterparty credit) risk. This risk premium was then clearly defined in a paper by Zhu and Pykhtin (\cite{ZP07}), under the name of Credit Value Adjustment ({CVA}). In the post-crisis era {CVA} became a key quantity to be taken into account  when trading derivatives in the OTC markets and this spurred a lot of research in the field:   see \cite{Gr12}, \cite{BMP13} and \cite{BCB} just to mention some.
In practice, {CVA} is  an adjustment of  the default-free value of a portfolio, to reduce this price in order to include the default risk. Along the years, other value adjustments have been introduced leading to the acronym {(X)VA}.

In the present paper we deal  only with plain vanilla (unilateral)  {CVA}. An important aspect of {CVA} and of its correct evaluation is the presence of Wrong Way Risk ({WWR}) that is, a decrease in the credit quality of the counterparty producing a higher exposure in the portfolio of the derivative's holder.  Under independence between  the exposure and the credit quality of the counterparty, computation of {CVA} simplifies, while it becomes computationally much more delicate if dependence is assumed.  To overcame this difficulty, several  methods have been proposed: Monte Carlo methods, from brute force to enhanced ones (see \cite{HW12} and \cite{RS12}), the copula method or {\it static} approach (see \cite{PR10}, \cite{C13}),  sharp bounding estimates (see \cite{GY16}). Here, we propose a new method and  we compare it with another  recently investigated in  \cite{BR18}.

In this paper we exploit the reduced-form or stochastic intensity approach, where the default event is characterized by means of a random time, representing the time of default, when the investor  might face either  a total loss or a partial  recovery of the investment's current value. Within this context, the computational difficulty in the evaluation of the {CVA} is twofold.  First, the default time might be not completely measurable with respect to the information generated by the market prices, since it reflects also other exogenous factors, secondly even under full knowledge of the default time, the derivative's evaluation calls  for the  joint distributions of the random time and the price processes,  usually very difficult to know.

In order to characterize the distribution of the default time, conditionally to the information generated by the market prices, under appropriate conditions the joint dynamics of the asset prices, of the default time and of the other stochastic factors can be described as a Markovian  system, whose components may  exhibit correlation. This correlation is going to be modeled by means of a set of parameters linking the processes driving the dynamics. In this framework, the usual theory of stochastic calculus allows to set up a PDE system, whose solution, though not easily computable, may be approximated. Several methods of approximations of PDE's are at disposal, the majority of them being based on some clever numerical discretization scheme, see e.g. \cite{KA03}.

In this paper we propose an alternative method, introduced in the papers \cite{AS} and \cite{ARS}, which expands theoretically the solution of the PDE system in a Taylor's series with respect to the correlation parameters.  Indeed, under quite general hypotheses, it is straightforward to verify that the solution to the PDE is regular with respect to the correlation parameters and therefore it can be  expanded in series around the zero value for all of them.The coefficients of the series are characterized, by using Duhamel's principle, as solutions to  a chain of PDE problems and they are therefore identified by means  of Feynman-Kac formulas and  expressed as expectations, that turn to be easier to compute or to approximate.

There are several  advantages in using  this method:
\begin{itemize}
\item Expanding  around the zero values of the correlation parameters means that the series coefficients are expectations of functionals of independent  driving processes, easier to compute or to approximate.
\item In many cases the zeroth term of the series can be explicit computed, increasing the precision of the approximation.
\item Comparing with Finite Differences methods or Monte Carlo methods, often a comparable accuracy is reached by the first order expansion.
\item Consequently the computational times are very little.
\item Compared to the other methods, our extends quite easily and in a straightforward way to multi-factor models, as shown is Section 5.
\end{itemize}

In the next section we introduce the general problem and setting,  in the third section we define our market model, while in the fourth we give the appropriate conditions for the convergence of the series and we show in  detail  how  the method works in absence of interest rate risk, finally a stochastic interest rate is considered  in the fifth section. It follows a short section recalling  the main features and results of the method \cite{BR18}, based on a change of measure technique and in the last section we provide numerical  comparisons among the different methodologies previously  discussed.

\section{CVA Evaluation of Vulnerable Options in an Intensity Model}

We consider   a finite time interval  $[0,T]$ and a complete probability space $(\Omega, \F, P)$, endowed with a filtration  $\{\Ft\}_{t\in [0,T]}$, augmented with the
$P-$null sets and made right continuous. We assume that all the processes have a c\'adl\'ag version.

The market is described by the interest rate process $r_t$ determining the money market account  denoted by
$B(t,s)= \mathrm e^{\int_t^s r_u du}$ and by a process  $X_t$  representing an asset log-price  (whose dynamics will be specified later), this process may depend also on other stochastic factors.  We assume

\begin{itemize}
\item that the filtration $\{\Ft\}_{t\in [0,T]}$ is rich enough to support all the aforementioned  processes;

\item to be  in absence of arbitrage;

\item that the given probability  $P$ is a risk neutral measure, already selected by some criterion.

\end{itemize}

In this market  a defaultable European contingent claim paying $f(X_T)$ at maturity is traded, where $f$ is a function whose  regularity properties will be specified later.
We denote by $\tau$ (not necessarily a stopping time w.r.t. the filtration $\F_t$) the default time  of the contingent claim and by   $Z_t$  an $\Ft-$measurable bounded recovery process.

To properly evaluate this type of derivative we need to include the information generated by the default time. We denote by $\G_t$ the progressively enlarged filtration, that makes $\tau$ a $\G_t-$stopping time, that is $\Gt=\F_t \lor \sigma(\{\tau\le t\})$.
From now on, we indicate by $H_t = \mathbf 1_{\{\tau \le t\}}$, the process generating the filtration $
\Ht$, so that $\Gt=\Ft\lor\Ht$.

 We make the fundamental assumption, known as the  H-hypothesis (see e.g. \cite{GJLR10} and \cite{G14} and the references therein), that

\medskip

\noindent
(H)\qquad \qquad\qquad \qquad  Every $\Ft-$martingale remains a $\Gt-$martingale.\hfill

\medskip \noindent
Under this assumption,  we may affirm that $\mathrm e^{X_s}/B(t,s)$ for $ s\ge t$ remains  a $\Gs-$martingale under the unique extension of the risk neutral probability to the filtration $\Gs$.
(To keep notation light, we do not indicate explicitly the probability we use for the  expectations, assuming that we are always working with the one corresponding to the filtration in use).

In this setting, for any given time $t\in [0,T]$,  the price of a defaultable claim, with positive final value  $f(X_T)$,  default time $\tau$ and recovery process $Z_t$, is given by
\begin{equation}
\label{eqvalue1}
c^d(t,T) = \E[B^{-1}(t,T) f(X_T) 1_{\{\tau > T\}} + B^{-1}(t,\tau) Z_{\tau} 1_{\{t < \tau \leq T\}} |\Gt],
\end{equation}
while the corresponding default free  value  is
\begin{equation}
\label{free}
c(t,T) = \E[B^{-1}(t,T)   f(X_T) |\Ft].
\end{equation}

Correspondingly the CVA, as a function of the running time and of the maturity, is given by
\begin{equation}
\label{CVA_0}
CVA(t,T) = \E(B^{-1}(t,\tau) Z_{\tau} 1_{\{t < \tau \leq T\}} |\Gt)=1_{\{\tau > t\}} [c(t,T) - c^d(t,T)].
\end{equation}
In many situations, investors do not know the default time and they may observe only whether it happened or not. The actual observable quantity is the asset price,  therefore  it  is interesting  to write the pricing formula \eqref{eqvalue1} in terms of $\F_t$, rather than in terms of $\Gt$. For that we have  the following Key Lemma, see \cite{BJR} or \cite{BCB}.

\begin{lemma}\label{key} For any integrable $\G-$measurable r.v. $Y$, the following equality holds
\begin{equation}
\label{key}
\E\Big[\mathbf 1_{\{\tau>
t\}}Y|\G_t\Big]=P(\tau>t|\Gt)\frac{\E\Big[\mathbf 1_{\{\tau>
t\}}Y|\Ft\Big]}{P(\tau>t|\Ft)}.
\end{equation}
\end{lemma}

Applying this lemma to the first and the second term of \eqref{eqvalue1} and recalling that $1-H_t=\mathbf 1_{\{\tau>t\}}$ is $\Gt-$measurable, we obtain
\begin{eqnarray}
\label{key1}
\E[B^{-1}(t,T) f(X_T) 1_{\{\tau > T\}}|\Gt]&=&\mathbf 1_{\{\tau>t\}}
\frac{\E[B^{-1}(t,T) f(X_T) 1_{\{\tau > T\}}|\Ft]}{P(\tau>t|\Ft)}\\
\label{key1.2}
\E[B^{-1}(t,\tau) Z_{\tau} 1_{\{t < \tau \leq T\}} |\Gt]&=&\mathbf 1_{\{\tau>t\}}\frac{\E[B^{-1}(t,\tau) Z_{\tau} 1_{\{t < \tau \leq T\}} |\Ft]}{P(\tau>t|\Ft)},
\end{eqnarray}
which  may be made more explicit by following the hazard process  approach.

We denote  the conditional distribution of the default time  $\tau$ given $\Ft$ by
\begin{eqnarray}
F_t=P(\tau\leq t|\Ft), \qquad \forall \, t\ge 0,
\end{eqnarray}
whence, for $u\ge t$,  $P(\tau\leq u|\Ft)=\E(P(\tau\leq u|\Fu)|\Ft)=\E(F_u|\Ft)$. If  $F_t(\omega)<1$ for all  $t>0$ (which automatically excludes that $\Gt\equiv \Ft$), we can  well define the so called $\F$- hazard process of $\tau$ as
\begin{equation}
\label{risk}
\Gamma_t:=-\ln(1-F_t)\quad\Rightarrow \quad F_t= 1 - \mathrm e^{-\Gamma_t}\quad \forall \, t>0, \qquad  \Gamma_0=0,
\end{equation}
moreover
\begin{equation}
\label{survival}
S_t:=1-F_t= \mathrm e^{-\Gamma_t}\quad \forall \, t>0, \qquad  S_0=1,
\end{equation}
is the $\F$-survival process. We assume $\Gamma_t$ to be differentiable. Its derivative, known as  the intensity process and denoted by $\lambda_t$, is such that
$\Gamma_t = \int_0^t \lambda_u du.$

Exploiting  \eqref{key1} and \eqref{key1.2} to pass to the $\cF_t$ filtration and  assuming that $B(t, \cdot)Z. $ is a bounded $\F-$ martingale (which is usually the case), by an extension of Proposition 5.1.1 of \cite{BR}, as developed in \cite{ARS1}, we may rewrite the pricing formula \eqref{eqvalue1} as
\begin{equation} \label{defaultprice}
\begin{aligned}
c^d(t,T) &= 1_{\{\tau > t\}} \E[\mathrm e^{-\int_t^T (r_s+\lambda_s) ds}f(X_T) | \Ft] \\
&+ 1_{\{\tau > t\}} \E[\int_t^T Z_{s} \lambda_s \mathrm e^{-\int_t^s (r_u+\lambda_u) du} ds| \Ft],
\end{aligned}
\end{equation}
recovering   formulas (3.1) and (3.3) in \cite{Lando98}, that the author obtained by modeling directly the random time $\tau$.

This formula can be specialized even further if we assume fractional recovery, $Z_t = R c(t,T)$  for some $0\le R< 1$. Using the Optional Projection Theorem, see e.g.  theorem 4.16 in \cite {N} , one gets to
\begin{equation}
\label{price2}
\begin{aligned}
c^d(t,T) &= 1_{\{\tau > t\}}\Big [ R \E[\mathrm e^{-\int_t^T r_udu} f(X_T) | \Ft]\\
&+ (1-R) \E[\mathrm e^{-\int_t^T (r_u+\lambda_u) du} f(X_T) | \Ft]\Big ],
\end{aligned}
\end{equation}
which can be interpreted as a convex combination of the default free price and the price with default.
As a consequence, from (\ref{CVA_0}) we have an expression also for the unilateral CVA as
\begin{equation}
\label{CVA_1}
CVA(t,T)=1_{\{\tau > t\}} (1-R) \E[\mathrm e^{-\int_t^T r_udu} f(X_T) (1-\mathrm e^{-\int_t^T \lambda_u du}) | \Ft].
\end{equation}

\begin{remark}
Last formula, by means of the survival process, could be briefly rewritten as
\begin{equation}
\label{CVA_survival}
CVA(t,T)=-1_{\{\tau > t\}} (1-R)\E[ \int_t^T \frac{f(X_T)}{B(t,T)}dS_u | \Ft].
\end{equation}
If   $G(t) = P(\tau > t) = \E[1_{\tau > t}]$  is  the (deterministic) survival function, assuming it  can be written as $G(t) = \e^{-\int_0^t h_s ds}$, for some  non-negative function   $h$, then we have that  $\E(S_t)=G(t)$ for all $t\geq 0$ (see (\ref{risk}) and (\ref{survival}))and
$$
dS_t = \lambda_t S_t dt=\frac{\lambda_t S_t}{h_t G(t)} dG(t) = \zeta_t dG(t)
$$
where we set  $\ds\zeta_t := \frac{\lambda_t S_t}{h_t G(t)}$.
Consequently, using the optional projection theorem,  the expectation in \eqref{CVA_survival} may be rewritten as
\begin{eqnarray*}
&&\E[ \int_t^T \frac{f(X_T)}{B(t,T)}dS_u | \Ft]= \E\Big [  \int_t^T \frac{f(X_T)}{B(t,T)}\zeta_u  dG(u)| \Ft\Big ]\\
&=&
 \E\Big [  \int_t^T \E [ \frac{f(X_T)}{B(t,T)}\zeta_u | \Fu  ] dG(u)| \Ft \Big ]=\E\Big [  \int_t^T \E [ \frac{f(X_T)}{B(t,T)} | \Fu  ] \zeta_u ]dG(u)| \Ft \Big ]\\
&=& \E\Big [  \int_t^T\frac{c(u,T) \zeta_u }{B(t,u)} dG(u)| \Ft \Big ]=\int_t^T\E[  \frac{c(u,T)  \zeta_u}{B(t,u)} | \Ft ] dG(u)
\end{eqnarray*}
and
\begin{equation}
\label{CVA_newformula}
CVA(t,T)=-1_{\{\tau > t\}} (1-R)\int_t^T\E[  \frac{c(u,T) \zeta_u}{B(t,u)} | \Ft ] dG(u).
\end{equation}
\end{remark}
For $t=0$  and a generic portfolio price process $V_t$ (the positive part $V_t^+$ coinciding in our case with $c(t,T)$, the default free price of the claim ) this formula is the starting point  of the analysis developed in  \cite{BR18}.

Finally we remark that
under independence between  $\lam_t$ and $(X_t,r_t)$, the second term  in \eqref{price2} simplifies further to
\begin{equation}
\label{ind2}
\E[\mathrm e^{-\int_t^T (r_s + \lam_s) ds}f(X_T)|\Ft] = \E[\mathrm e^{-\int_t^T r_s ds}f(X_T)|\Ft] \E[\mathrm e^{-\int_t^T \lam_s ds}|\Ft].
\end{equation}
Correspondingly, we get a similar factorization for the CVA 
\begin{eqnarray} \label{CVA_independent}
CVA(t,T) & = &1_{\{\tau > t\}} (1-R) \E[\mathrm e^{-\int_t^T r_udu} f(X_T) | \Ft]  \E[(1-\mathrm e^{-\int_t^T \lambda_u du}) | \Ft] \nonumber \\
 & = & 1_{\{\tau > t\}} (1-R) c(t,T) \frac{P(t < \tau \leq T| \Ft)}{P(\tau \geq t | \Ft)},
\end{eqnarray}
where the last equality follows from the Key Lemma and the definition of hazard process (see e.g. \cite{BR}, Sect. 8.2). In this case, the two factors are respectively the price of a European derivative and the price of a bond.
Thus we may arrive at explicit formulas whenever the models for $X$ and $\lambda$ are appropriately chosen.

\section{The model}

We assume that in the given probability space, the following  diffusion dynamics are satisfied
\begin{eqnarray}
\label{SDE1}
X_s &=& x + \int_t^s (r_u-\frac{\sigma^2}2)du + \sigma  (B_s-B_t), \qquad x\in\mathbb R \\
\label{SDE4} \lam_s&= &
\lam + \int_t^s \gamma(\theta - \lam_u) du +\eta \int_t^s \sqrt{\lam_u}dY_u,\quad \lam>0 \\
\label{SDE3}r_s& =& r  + \int_t^s k(\mu-r_u)du + \nu (W_s-W_t),\qquad r>0,
\end{eqnarray}
where  the parameters  are such that $k , \theta,\eta, \sigma, \mu>0$, $\gamma, \nu\ge 0$, $2k\theta>\eta^2$ and $B , Y, W$ are correlated Brownian motions with a given correlation matrix.  To simplify calculations, in what follows we assume independence between the interest rate and default intensity, i.e. between $Y$ and $W$; with this choice we may represent the triple  $B, Z, W$ as
$$
B_t = \rho B^1_t+ \delta B^2_t + \sqrt{1-\rho^2- \delta^2} B^3_t, \ \ \ Y_t = B^1_t,\ \ \ W_t = B^2_t;
$$
where $(B^1,B^2,B^3)$ is a 3-dimensional Brownian motion and $\delta^2+\rho^2 \le 1$.

\medskip
We remark that under independence we have an explicit expression of the factor  $\ds E[\mathrm e^{-\int_t^T \lam_s ds}|\Ft]$ appearing in \eqref{ind2}, being the bond price with a CIR interest rate. The problem is then reduced to computing the other factor representing the price of the European derivative.

\section{Correlation expansion}
\label{our}

For the sake of simplicity, in this section we assume $R=0$ and the short rate to be constant, $r_t \equiv r$ for all $t\in [0,T]$. To consider $r$ a function in time is a straightforward generalization, while a stochastic interest rate will be considered specifically in the next section.

The model, which we write in flow notation, is hence reduced to
\begin{equation}
\label{SDEsystem}
\begin{cases}
X_s^{t,x,\lambda} =x + (r- \frac{\sigma^2}{2})(s\!-\!t)+ \sigma \Big [\rho (B_s^1\!-B^1_t)+\sqrt{1\!-\!\rho^2}(B^2_s-B_t^2)\Big ]\\
\lam_s^{t,\lambda} = \lambda + \int_t^s\gamma(\theta-\lam_u^{t,\lambda})du+\int_t^s \eta\sqrt{\lam_u^{t,\lambda}}dB^1_u.
\end{cases}
\end{equation}
The two-dimensional diffusion $\mathbf {U}^{t,x,\lam,\rho}_t:=(X_s^{t,x,\lam},\lam_s^{t,\lam})$ is a Markov process since the coefficients,
$$
\mu(x, \lam):=
\left( \begin{array}{c}
r- \frac{\sigma^2}{2} \\
\gamma(\theta-\lam)
\end{array} \right), \quad \textrm{and } \quad
\Sigma(x,\lam ):= \left( \begin{array}{cc}
\sigma\rho & \sigma\sqrt{1-\rho^2}\\
  \eta\sqrt{\lam}& 0 \\
\end{array} \right)
$$
are deterministic. This implies that the price $c^d(t,T)$ of any European defaultable derivative with payoff
$F(\mathbf {U}^{t,x,\lam,\rho}_T)$ will be a deterministic function $u(\cdot)$  of all the initial data, that is
\begin{equation}\label{basic_expect}
u(x,\lam,t,T;\rho)=\mathrm e^{-r(T-t)}\E(\e^{-\int_t^T\lam^{t,\lam}_s ds}F(\mathbf {U}^{t,x,\lam,\rho}_T)).
\end{equation}
We remind that this computation is a  crucial step towards the evaluation of the defaultable derivative  \eqref{price2} and of the corresponding CVA.

When $\rho=0$, the vector process $\mathbf {U}^{t,x,\lam,0}_t$  is also affine, since both
$$
\mu(x, \lam) \qquad  \textrm { and } \quad
\Sigma(x, \lam)\Sigma(x,\lam)'= \left( \begin{array}{cc}
\sigma^2  & 0 \\
0& \eta^2\lam
\end{array} \right)
$$
have components which are affine functions. Therefore one may employ Fourier   transform techniques to evaluate  $u(x,\lam,t;0)$. In particular, if  the payoff $F$ is chosen in the class of affine functions
$$
F(\mathbf {U}^{t,x,\lam,0}_T)=\e^{\mathbf {v}\cdot\mathbf {U}^{t,x,\lam,0}_T},\quad\quad\;\; \mathbf  {v}=(v_1,v_2)\in \mathbb C^2,
$$
then also the conditional expectation is  exponentially affine
$$
u(x,\lam,t,T;0)=\e^{-r(T-t)+ \alpha(T-t)\cdot \mathbf {U}^{t,x,\lam,0}_t},
$$
for some complex-valued vector function $\alpha(s)=(\alpha_1,(s) \alpha_2(s))$, whose components verify a Riccati system of ODE's with initial values $(\alpha_1(0),\alpha_2(0))=(v_1,v_2)$. In the present paper we shall consider the payoff of a plain vanilla call option written on the stock, hence
$\ds
F(\mathbf {U}^{t,x,\lam,\rho}_T)=f(X_T):=(\e^{X_T}-K)^+,
$
but our methodology may be extended to other derivatives. For $\rho=0$, even if the payoff is not exponentially affine (unless $K=0$), it is possible to reduce the problem to that case and solve it by Fourier transform.

When $\rho\neq 0$, the power of Fourier transform is lost and we have to resort to alternative method to evaluate \eqref{basic_expect}.

Here we use a technique introduced in \cite {AS} and \cite{ARS}, that gives an expression of $u(x,\lam,t;\rho)$ as a power series of $\rho$ around 0
$$
u(x,\lam,t,T;\rho)=\sum_{k=0}^{\infty}\frac{\partial^ku}{\partial\rho^k}\big|_{\rho=0}\frac{\rho^k}{k!},
$$
since it is quite immediate to show that $u$ depends smoothly on the correlation parameter.
Since the diffusion coefficient of $X$ is a constant, $\sigma >0$, the conditions in \cite {AS} to guarantee  this power series has a strictly positive convergence radius are automatically satisfied, as long as $F$ is an integrable payoff.

The series expansion gives a tool to approximate $u(x,\lam,t,T;\rho)$, by stopping it at any chosen order.
The coefficient $g_0(x,\lam,t,T)$  equals $u(x,\lam,t,T;0)$ and it can be computed in closed form. As we mentioned before, this corresponds to the independent case when the vector process $\mathbf U$ is affine. All  the other  coefficients, $g_k(x,\lam,t,T)$ can be iteratively computed  by  exploiting the Duhamel's principle, as we are going to show.

By the Feymann-Kac formulas, $u(x,\lam,t,T;\rho)$   solves the parabolic PDE
\begin{equation}\label{pde1}
\begin{cases}
 \ds&\frac{\partial u}{\partial t}+{\cL^{\rho}}u=0\\
&u(x,\lam,T,T;\rho)=(\e^x-K)^+,
\end{cases}
\end{equation}
where we denoted $  \cL^{\rho}= \cL^0+\rho \cA$, with
\begin{eqnarray}
\label{op0}
 \cL^0&:=&\frac{\sigma^2}{2}\frac{\partial^2}{\partial x^2}+\frac{\eta^2\lam}{2}\frac{\partial^2}{\partial \lam^2}+(r-\frac{\sigma^2}{2})\frac{\partial }{\partial x}+\gamma(\theta-\lam)\frac{\partial}{\partial \lam}-r-\lam\\
 \label{opmix}
\cA &:=& \eta\sigma\sqrt{\lam}\frac{\partial^2}{\partial x\partial \lam}.
\end{eqnarray}
By differentiating and taking $\rho=0$, it is readily seen that  the coefficients $g_k(x,\lam,t,T)$ must respectively satisfy the following parabolic equations
\begin{equation}\label{g0}
\begin{cases}
&\frac{\partial g_0}{\partial t}+ \cL^0g_0=0\\
& g_0(x,\lam,T,T)=(\e^x-K)^+,
\end{cases}\quad
\begin{cases}
&\frac{\partial g_k}{\partial t}+ \cL^0 g_k=-\cA g_{k-1}\\
&g_k(x,\lam,T,T)=0.
\end{cases}\quad k\ge 1.
\end{equation}
Once again, by the Markov property and Feymann-Kac formulas, $g_0(x,\lam,t,T)$ admits the following representation
\begin{eqnarray} \label{g0_explicit}
g_0(x,\lam,t,T)&=&\e^{-r(T-t)}\E(\e^{-\int_t^T\lam^{t,\lam}_s ds}(\e^{X_T^{t,x,\lam}}-K)^+) \nonumber \\
&=&\E(\e^{-\int_t^T\lam^{t,\lam}_s ds})\e^{-r(T-t)}\E((\e^{X_T^{t,x}}-K)^+),
\end{eqnarray}
where in the last passage we used the independence of the processes $(X_t)$ and $(\lam_t)$ ($\rho=0$).
The first factor is the bond price with a CIR process and presents an exponentially affine solution, while the second is the usual Black \& Scholes  price of a European call option, $c_{BS}(x,t,T)$, hence we have
\begin{equation}
\label{expl}
\begin{aligned}
&g_0(x,\lam,t,T)=\e^{-\alpha_1(T-t)-\alpha_2(T-t)\lam}c_{BS}(x,t,T)\\
=&\e^{-B_1(T-t)-B_2(T-t)\lam}\big[\e^xN(d_1(x,T-t)-K\e^{-r(T-t)}N(d_2(x,T-t))\big],
\end{aligned}
\end{equation}
where $d_{1,2}(x,\tau)= \frac{x-\kappa + (r\pm\frac {\sigma^2}2)\tau}{\sigma\sqrt{\tau}}$ , $ \kappa= \ln K$ and
\begin{eqnarray}
\label{bondprices}
B_1 (\tau)&=&\frac {2\gamma \theta}{\eta^2}\ln\left (\frac{2\beta\e^{\frac{\gamma+\beta}2\tau}}{\beta -\gamma+ (\gamma+\beta)\e^{\beta \tau}}\right )\\
B_2(\tau)&= &\frac{2(\e^{\beta \tau }-1)}{\beta-\gamma+ (\gamma+\beta)\e^{\beta \tau}},
\end{eqnarray}
with $\beta=\sqrt{\gamma^2+\eta^2}$, $\tau=T-t$.
The other  equations of \eqref{g0} can be solved by Duhamel's principle which states that
$$
g_k(x,\lam,t,T)=-\int_t^Tg_k^{\alpha}(x,\lam,t)d\alpha,
$$
where  $g_k^{\alpha}(x,\lam,t)$  is the solution to the PDE problem
\begin{equation}
\label{PDE1}
\begin{cases}
&\frac{\partial g^{\alpha}_k}{\partial t}+ {L^{0}}g^{\alpha}_k=0,\\
&g^{\alpha}_k(x,\lam,\alpha)=-\cA g_{k-1}(x,\lam,\alpha,T)
\end{cases}
\end{equation}
for any fixed $\alpha\in(t, T]$. This sets up an iterative procedure to compute theoretically the coefficients of any order, by means of a repeated application of Feymann-Kac formulas. Indeed we have
\begin{eqnarray*}
g_k(x,\lam,t,T)&=&-\int_t^Tg_k^{\alpha_k}(x,\lam,t)d\alpha_k\\
&=&
\int_t^T\E\Big(\e^{-r(\alpha_k-t)}\e^{-\int_t^{\alpha_k}\lam^{t,\lam}_s ds}\cA g_{k-1}(X^{t,x}_{\alpha_k},\lam^{t,\lam}_{\alpha_k},\alpha_k,T)\Big)d\alpha_k
\end{eqnarray*}
and we can iterate the procedure arriving to a formula involving $k$ integrals but depending only on $g_0$.

Inevitably, coefficients of higher order are harder to compute. In the hope to obtain good numerical results, we consider  the first order approximation
$$
\bar{u}(x,\lam,t,T;\rho):=u(x,\lam,t,T;0)+(\frac{\partial u}{\partial\rho}\big|_{\rho=0})\rho\equiv g_0(x,\lam,t,T)+g_1(x,\lam,t,T)\rho.
$$

From\eqref{expl}, we may explicitly compute
\begin{eqnarray*}
\cA g_0(x,\lam,t,T)&=&\eta\sigma\sqrt{\lam}\frac{\partial^2}{\partial x\partial \lam}g_0(x,\lam,t,T)\\
&=&-\eta\sigma\sqrt{\lam}B_2(T-t)\e^{-B_1(T-t)-B_2(T-t)\lam}\frac{\partial}{\partial x}c_{BS}(x,t,T)\\
&=&-\eta\sigma\sqrt{\lam}B_2(T-t)\e^{-B_1(T-t)-B_2(T-t)\lam}\e^x N(d_1(x,T-t))
\end{eqnarray*}
and consequently
\begin{eqnarray*}
\!\!\!&&\!\!g_1(x,\lam,t,T)=-\int_t^Tg_1^{\alpha}(x,\lam,t)d\alpha\\
\!\!\!&=&\!\!\int_t^T\E\Big(\e^{-r(\alpha-t)}\e^{-\int_t^{\alpha}\lam^{t,\lam}_s ds}\cA g_0(X^{t,x}_{\alpha},\lam^{t,\lam}_{\alpha},\alpha,T)\Big)d\alpha\\
\!\!\!&=&\!\!-\eta\sigma\! \int_t^T\!\! \!\E\Big[\sqrt{\lam^{t,\lam}_{\alpha}}B_2(T\!-\!\alpha)\e^{-B_1(T\!-\!\alpha)-B_2(T\!-\!\alpha)\lam^{t,\lam}_{\alpha}}\!\e^{X^{t,x}_{\alpha}}\! N(d_1(X^{t,x}_{\alpha}\!,T\!-\alpha))\Big]d\alpha
\end{eqnarray*}
We remark that the expectation in  the integral  is to be evaluated under independence of the two processes $X$ and $\lam$, therefore we have
\begin{equation}
\label{g1alpha}
\begin{aligned}
\!\!\!&\eta \sigma\E\Big(\e^{-r(\alpha-t)}\e^{-\int_t^{\alpha}\lam^{t,\lam}_s ds}\cA g_0(X^{t,x}_{\alpha},\lam^{t,\lam}_{\alpha},\alpha,T)\Big)\\
\!\!\!=&\Gamma(t,\alpha,T)\E\Big[\sqrt{\lam^{t,\lam}_{\alpha}}\e^{-B_2(T\!-\alpha)\lam^{t,\lam}_{\alpha}-\int_t^{\alpha}\lam^{t,\lam}_s ds}\Big ] \E\Big [\e^{X^{t,x}_{\alpha}} \!N(d_1(X^{t,x}_{\alpha},T\!-\alpha))\Big]
\end{aligned}
\end{equation}
where $\Gamma(t,\alpha,T)\equiv \eta\sigma e^{-r(\alpha-t)}B_2(T-\alpha)\e^{-B_1(T-\alpha)}$.

From the above formula  we remark that $g_1(x,\lam,t)<0$ implying that  the price of the defaultable European call increases with $\rho$ in a small interval around $\rho=0$.
\begin{remark} \label{WWR_g1} Let $t=0$ , then from (\ref{CVA_0}) and (\ref{g0_explicit}) we have
\begin{equation}\label{linear}
\begin{aligned}
CVA(0,T)  = & c(0,T) - c^d(0,T)\\
 \approx &c(0,T) - g_0(x,\lam,0,T) - g_1(x,\lam,0,T) \rho \\
=& c(0,T) - c(0,T)P(\tau > T)  - g_1(x,\lam,0,T) \rho\\
=& c(0,T) P(\tau \leq T) - g_1(x,\lam,0,T) \rho.
\end{aligned}
\end{equation}
The first term on the right-hand side represents the CVA under independence between the default event and the exposure (see (\ref{CVA_independent})). Hence $g_1(x,\lam,0,T)$ measures the impact of the factor correlation  on CVA.
\end{remark}
We now focus on the  first expectation in \eqref{g1alpha}. Let us set $b_\alpha:=B_2(T-\alpha)$  let us  condition internally with respect to $\lam^{t,\lam}_{\alpha}$, obtaining
\begin{eqnarray*}
&&\E\Big[\sqrt{\lam^{t,\lam}_{\alpha}}\e^{-b_\alpha\lam^{t,\lam}_{\alpha}-\int_t^{\alpha}\lam^{t,\lam}_s ds}\Big ]\\
&=& \int_0^{+\infty}\E\Big [ \sqrt{\lam^{t,\lam}_{\alpha}}\e^{-b_\alpha\lam^{t,\lam}_{\alpha}-\int_t^{\alpha}\lam^{t,\lam}_s ds}| \lam^{t,\lam}_{\alpha}=\zeta\Big  ]f_{\lam^{t,\lam}_{\alpha}}(\zeta) d\zeta\\
&=& \int_0^{+\infty}\sqrt\zeta\e^{-b_\alpha\zeta}E\Big [\e^{-\int_t^{\alpha}\lam^{t,\lam}_s ds}| \lam^{t,\lam}_{\alpha}=\zeta\Big  ]f_{\lam^{t,\lam}_{\alpha}}(\zeta) d\zeta.
\end{eqnarray*}
The density  $f_{\lam^{t,\lam}_{\alpha}}$ is explicitly known ( see for instance \cite{AL}). Moreover in \cite{LLPB} or in \cite{PP} an explicit expression of the conditional moment generating function of $\ds \int_t^{\alpha}\lam^{t,\lam}_s ds$ is provided as
$$
E\Big [\e^{-\int_t^{\alpha}\lam^{t,\lam}_s ds}| \lam^{t,\lam}_{\alpha}=\zeta\Big  ]= \frac{M_{t,\alpha}(\lam,\zeta)}{f_{\lam^{t,\lam}_{\alpha}}(\zeta)}I_{\nu}\Big(\frac{2\bar{\gamma}\sqrt{\zeta\lam}}{\sigma^2\sinh\big(\frac{\bar{\gamma}(\alpha-t)}{2}\big)}\Big),
$$
where $\nu= \frac{2\gamma\theta}{\sigma^2}-1$, $\bar{\gamma}= \sqrt{\gamma^2+2\sigma^2}$,
$$
I_{\nu}(z)\equiv (\frac{z}{2})^{\nu}\sum_{n=0}^{\infty}\frac{(\frac{z^2}{4})^n}{n!\Gamma(\nu+k+1)}
$$
is the modified Bessel function of the first kind and
$$
M_{t,\alpha}(\lam,\zeta)=\frac{2\bar{\gamma}}{\sigma^2}\big(\frac{\zeta}{\lam}\big)^{\frac{\nu}{2}}
\frac{\e^{-\frac{\bar{\gamma}(\alpha-t)}{2}-\frac{1}{\sigma^2}[\bar{\gamma}(\lam+\zeta)\frac{\e^{\bar{\gamma}(\alpha-t)}+1}{\e^{\bar{\gamma}(\alpha-t)}-1}-\gamma(\lam-\zeta)-\theta\gamma^2(\alpha-t)]}}{1-\e^{-\bar{\gamma}(\alpha-t)}}.
$$
Setting $a_{n}(\nu)\equiv [2^{\nu+2n} n!\Gamma(\nu+n+1)]^{-1}$ and
$
z_{t,\alpha}(\lam,\zeta)=\frac{2\bar{\gamma}\sqrt{\zeta\lam}}{\sigma^2\sinh\big(\frac{\bar{\gamma}(\alpha-t)}{2}\big)}
$, we may write our expectation as a power series
$$
\E\Big[\sqrt{\lam^{t,\lam}_{\alpha}}\e^{-b_\alpha\lam^{t,\lam}_{\alpha}-\int_t^{\alpha}\lam^{t,\lam}_s ds}\Big ]
\!=\!\sum_{n=0}^{\infty}a_n(\nu)\!\!\int_0^{+\infty}\!\!\!\!\!\!\sqrt\zeta\e^{-b_\alpha\zeta}M_{t,\alpha}(\lam,\zeta)
[z_{t,\alpha}(\lam,\zeta)]^{\nu+2n}d\zeta
$$
that can be truncated at any given order.

Since $X_{\alpha}^{t,x} \sim N(x+(r-\frac{\sigma^2}{2})(\alpha-t)),\sigma^2(\alpha-t))$, the second expectation in  \eqref{g1alpha} becomes

$$
\E\Big [\e^{X^{t,x}_{\alpha}} \!N(d_1(X^{t,x}_{\alpha},T\!-\alpha))\Big]\!=\!\!
\int_{\mathbb R} \!\!\e^y
N(d_1(y,T-\alpha)) \frac {
\exp\left \{\frac{ [y-x-(r-\frac{\sigma^2}2)(\alpha-t)]^2}{\sigma^2(\alpha-t)}\right\}}{\sqrt{ 2\pi \sigma^2(\alpha-t)}} dy.
$$

\section{A three-factor model}
\label{threefac}
In this section we shortly present the correlation expansion for the more general market model (\ref{SDE1}),  to show that the method can be easily extended to multi-factor models. Indeed  the methodology remains the same and it is just a matter of handling slightly more complex calculations that lead nevertheless to computable formulas. As in the previous section we take $R=0$.

Let $\cc=(\rho,\delta)$ be the correlations vector, then by the Feymann-Kac theorem, the call price $u(x,\lam,r,t,T;\cc)$ must solve the following parabolic PDE:
\begin{equation}
\label{PDEtwo_fact}
\left\{ \begin{array}{l}
          \frac{\partial u}{\partial t}+ L^{\cc}u=0 \\
          u(x,\lam,r, T,T;\cc)=(\e^{X_T}-K)^+
        \end{array}
\right.
\end{equation}
where
$$
 {L^{\cc}}\equiv {L^0}+\rho(\sigma \eta \sqrt{\lam} \frac{\partial^2}{\partial x\partial \lam})+
 \delta (\sigma \nu \frac{\partial^2}{\partial x \partial r}) \equiv  {L^0}+{\cc}\cdot (A_{\rho}, A_{\delta})
$$
and
$$
 {L^0}\equiv \frac{\sigma^2}{2}\frac{\partial^2}{\partial x^2}+\frac{\eta^2\lam}{2}\frac{\partial^2}{\partial \lam^2} + \frac{\nu^2}{2}\frac{\partial^2}{\partial r^2}+(r-\frac{\sigma^2}{2})\frac{\partial }{\partial x}+\gamma(\theta-\lam)\frac{\partial}{\partial \lam}+ k(\mu-r)\frac{\partial}{\partial r}-r-\lam
$$
By definition the first-order approximation of the call price is given by
\begin{equation}
\label{approxthree_fact}
\bar u(x,\lambda,r,t,T;\cc) \equiv g_0(x,\lambda,r,t,T) + {\cc} \cdot \underline{g}_1(x,\lambda,r,t,T)
\end{equation}
where $g_0$ solves (\ref{PDEtwo_fact}) with ${\cc}=(0,0)$ and $\underline{g}_1 = (v, w)'$. The functions $v=v(x,\lambda,r,t,T)$ and $w=w(x,\lambda,r,t,T)$ can be computed by the same method used in section 4 as we are showing below. Indeed by the Feymann-Kac theorem and the independence of the processes at $\cc=\mathbf{0}$, we first get explicitly $g_0$ as
\begin{eqnarray*}
g_0(x,\lam,r,t,T) & = & \E(\e^{-\int_t^T\lam_s^{t,\lam} ds})\E(\e^{-\int_t^T r_s^{t,r} ds}\e^{X_T^{t,x}}-K)^+) \\
 & = & \e^{-B_1(T-t)-B_2(T-t)\lam}c_{BS}^V(x,r,t,T),
\end{eqnarray*}
where $c_{BS}^V(x,r,t,T) = \e^x N(D_1) - K P^r(r,t,T) N(D_2)$. Here $P^r(r,t,T) = \e^{-A_1(T-t) - A_2(T-t) r}$ is the Vasicek ZCB price maturing at $T$ and the functions $D_{1,2}=D_{1,2}(x,r,V(T-t))$  and $V(T-t)$ are known (see \cite{Rab89}). Then the derivatives $\ds\frac{\partial}{\partial x} c_{BS}^V$ and $\ds\frac{\partial}{\partial r} c_{BS}^V$ are also explicitly computable and so are the terms $A_{\rho} g_0$ and $A_{\delta} g_0$. By Duhamel's principle we get
$$
v(x,\lambda,r,t,T)=\int_t^T \!\!\! v^{\alpha}(x,\lambda,r,t,T) d\alpha, \ \ \ w(x,\lambda,r,t,T)=\int_t^T \!\!\! w^{\alpha}(x,\lambda,r,t,T) d\alpha,
$$
where $v^{\alpha}$ and $w^{\alpha}$ solve PDE's analogous to   (\ref{PDE1}). They are given by:
\begin{eqnarray*}
v^{\alpha}(x,\lambda,r,t,T)=&& \sigma \eta \mathrm B_2(T-\alpha)\mathrm e^{-B_1(T-\alpha)} \E\Big[\sqrt{\lambda_{\alpha}^{t,\lam}}\mathrm e^{-\int_t^{\alpha}\lambda_s^{t,\lam} ds-B_2(T-\alpha)\lambda_{\alpha}^{t,\lam}}\Big]\\
&&\times \E\Big[\mathrm e^{X_{\alpha}^{t,x}} N(D_1(X_{\alpha}^{t,x},r_{\alpha}^{t,r},\sigma,T-\alpha))\Big]
\end{eqnarray*}
(notice that all processes are evaluated for  ${\cc}=(0,0)$) and
\begin{eqnarray*}
w^{\alpha}(x,\lambda,r,t,T)=&& - \sigma \nu \frac{A_2(T-\alpha)}{\sqrt{V(T-\alpha)}}
\E\Big[\e^{-\int_t^{\alpha}\lambda_s^{t,\lam} ds-B_2(T-\alpha)\lambda_{\alpha}^{t,\lam}}\Big ] \\
&& \times \E\Big[ \e^{-\int_t^{\alpha}r_s^{t,r} ds} \e^{X_{\alpha}^{t,x}} N'(D_1(X_{\alpha}^{t,x},r_{\alpha}^{t,r},V(T-\alpha)))\Big].
\end{eqnarray*}
The expectations involving only the intensity process are similar to those of the previous section. The other expectations are relative to Gaussian processes. Therefore  (\ref {approxthree_fact}) is numerically fully implementable.

\section{CVA  and the change of measure approach}
\label{Brins}

Recently Brigo and Vrins \cite{BR18} proposed a method for addressing the CVA computational problem under WWR based on a change of measures, e.g. Girsanov's theorem, in the stochastic-intensity default setup. Their starting point is the following  formula for the time-zero CVA (compare with (\ref{CVA_newformula})) of portfolio price process $V_t$ :
\begin{equation}
\label{CVA_Brigo}
CVA(0,T) = -(1-R) \int_0^T \E[\frac{V_t^+}{B(0,t)} \zeta_t] dG(t),
\end{equation}
where $\E[\cdot]$ is the expectation under the risk-neutral measure. The $EPE$ (expected positive exposure) under WWR is the function
$$
EPE(t) = \E[\frac{V_t^+}{B(0,t)} \zeta_t].
$$
Girsanov's theorem is used to factorize the EPE. Indeed
by defining an equivalent martingale measure $Q^{C^{\F,t}} \sim Q$ as
$$
Z_s^t := \frac{dQ^{C^{\F,t}}}{dQ} = \frac{M^t_s}{M^t_0}, \quad\textrm{where } M_s^t = \E[\frac{1}{B(0,t)} \lambda_t S_t | \cF_s], \ s \in [0,t],
$$
in \cite{BR18} they prove that
$$
\E[\frac{V_t^+}{B(0,t)} \zeta_t] = \E^{C^{\F,t}}[V_t^+] \E[\frac{ \zeta_t}{B(0,t)}].
$$
The measure $Q^{C^{\F,t}}$ is called wrong-way measure and it is associated to the num\'{e}raire $C^{\F,t}_{\cdot} = B(0,\cdot) M_{\cdot}^t$.



In order to apply such a methodology, it is therefore necessary to obtain the dynamics of $V_t$ under the measure $Q^{C^{\F,t}}$. By assuming a continuous dynamic for $V_t$ under $Q$ described by a SDE, the change of measure results in a drift adjustment, we refer to \cite{BR18} for the full details.

In \cite{BRH18} Brigo et al. applied the results obtained in \cite{BR18}   to the calculation of CVA under WWR for a call option in the market model described by  (\ref{SDEsystem}). The risk free rate being constant implies that  $\E[B(0,t)^{-1}\zeta_t]=-\e^{-rt}$. Moreover the explicit expression of the new drift is
\begin{equation}
\label{newdrift}
\theta_t^s \equiv \theta_t^s(\lam_t) = \rho \eta \sqrt{\lam_t} \left(\frac{A^{\lambda}(s,t) B_t^{\lambda}(s,t)}{A^{\lambda}(s,t) B_t^{\lambda}(s,t) \lambda_t - A_t^{\lambda}(s,t)} - B^{\lambda}(s,t) \right),
\end{equation}
the functions $\log A^{\lambda}=-B_1$ and $B^{\lambda}=B_2$ being as in (\ref{bondprices}).
In order to be able to compute the expectations, it was necessary to replace the process $\lambda_t$ with a deterministic proxy $\lambda(t)$ in (\ref{newdrift}).
Once the chosen approximant is plugged into (\ref{newdrift}), the expression $EPE(t) = -\e^{-rt}\E^{C^{\F,t}}[c(t,T)]$ can be evaluated analytically leading to (see \cite{BRH18})
\begin{equation} \label{integrand_CVA_Brigo}
\begin{aligned}&E^{C^{\F,t}}[\frac{c(t,T)}{B(0,t)}] \\
\approx &\e^{x_0+\sigma \Theta_t} N\left(\frac{\hat\alpha(t)+\beta(t) \sigma \sqrt{t}}{\sqrt{1+\beta^2(t)}} \right) - \e^{\kappa-rT} N\!\left(\frac{\hat\alpha(t)- \sigma \sqrt{T-t}}{\sqrt{1+\beta^2(t)}} \right)
\end{aligned}
\end{equation}
where
\begin{eqnarray*}
\Theta(t)&=& \int_0^t \theta(u,t)du,\quad \theta(u,t)=\theta^t_u(\lam(u)),
\quad
\hat \alpha(t )= \alpha(t) + \frac{\Theta_t}{\sqrt{T-t}}\\
\alpha(t) &=& \frac{1}{\sigma \sqrt{T-t}} \left(x_0-\kappa + \left(r+ \frac{\sigma^2}{2} \right)T - \sigma^2 t\right), \quad \beta(t) = \sqrt{\frac{t}{T-t}}.
\end{eqnarray*}
Two deterministic proxies $ \lambda(t)$ were considered: $\E[\lambda_t]$ and $\E^{C^{\F,t}}[\lambda_t]$. While the first is analytically known, the second requires a further approximation step (see \cite{BRH18}). Inserting (\ref{integrand_CVA_Brigo}) in (\ref{CVA_Brigo}) a numerical integration procedure gives the CVA under WWR.

\begin{remark}\label{Kim}
It should be noticed that other methods based on the approximation of the process $(\lam_t)$  could be exploited in order to price a vulnerable call option in the market model (\ref{SDEsystem}), and hence its CVA. For instance,  the volatility expansion method of Kim and Kunimoto, see \cite{KK99}, considers  a Taylor expansion of the process  $(\lam_t)$ in powers of $\eta$ around $\eta=0$. Stopping the series at the first order in $\eta$ and setting $\lam(s)=\lam \exp(-\gamma(s-t))+\theta(1- \exp(-\gamma(s-t)))$, they have for all $s\geq t$ and $\lam_t=\lam$:
\begin{equation}
\label{KK1}
\lam_s=\lam(s)+\eta \int_t^s \e^{-\gamma(s-u)}\sqrt{\lam(u)}(\rho dB^1_u+\sqrt{1-\rho^2}dB^2_u)+o(\eta).
\end{equation}
Inserting the approximation (\ref{KK1}) in the evaluation formula for the vulnerable call option, after some manipulations the following result is obtained
\begin{equation}
\label{KK2}\!\!
u(x,\lam,t,T;\rho)\approx \e^{-\int_t^T\lam(s)ds}\big[c_{BS}(x,t,T)\!-\!\rho\sigma\eta \e^{x-\frac{\sigma^2}{2}(T\!-t)}N(d_1)\Lambda(\lam,t,T)]
\end{equation}
with $c_{BS}$ denoting the classical Black-Scholes price and
$$
\Lambda(\lam,t,T)=\int_t^T \int_u^T \e^{-\gamma(s-u)}\sqrt{\lam(u)}duds.
$$
\end{remark}
In the next section we are going to provide  a comparison of the numerical performances of the different methods which have been presented.

\section{Numerical results}

In this section we compare numerically our method to compute the CVA for a vulnerable option with the methods mentioned above, exploiting the Monte Carlo approximations as a benchmark.

We considered model (\ref{SDEsystem}) with exogenously chosen parameters  $\gamma=0.2$, $\theta=0.05$, $\lambda_0=0.04$ 
and $S_0=100$.
Instead, we varied $\rho$, $\sigma$ and $\eta$ to check the performances of the methods. Positive correlation values relate to the WWR effect on the call option. The strike price is fixed to $K=100$ and the maturity is $T=1$: without loss of generality we also set the risk-free rate $r=0$ and $t=0$. All the pricing methods have been implemented in MatLab (R2017).

For the benchmark, Monte Carlo method was implemented with an Euler discretization of the CIR process, while the geometric Brownian motion was exactly simulated. In order to improve the Monte Carlo estimates, we implemented a control variate technique by using the default-free call price as a control. In these experiments we set $n=1000$ time step points in $[0,T]$ and $M=1\, 000 \, 000$ samples.

For the first order approximation of the expansion we proposed in section \ref{our},  we computed $g_0$ analitically, while for $g_1$,  we first computed the term $g_1^{\alpha}$ on a grid of equispaced points $\alpha_k$ in $[0,T]$ by using the adaptive Gauss-Kronrod quadrature algorithm and  then the resulting vector was interpolated and finally integrated by using once again the GK algorithm to get $g_1$. On a Intel Core i7 (\@2.40 GHz), the whole procedure requires about $0.3$ secs. Of course, the CVA approximation for different values of $\rho$ is simply obtained by linearity, see eq. (\ref{linear}), without any further computational cost.

The drift adjustment method recalled in section \ref{Brins} is based on the replacement of the process $\lambda_t$ with a deterministic proxy in the drift (\ref{newdrift}). As it was pointed out, different choices can be made: we have chosen to implement  $\lambda(t)=\E[\lambda_t]$. Inserting (\ref{integrand_CVA_Brigo}) in (\ref{CVA_Brigo}) a numerical integration procedure gives the CVA. This numerical approximation (taking about $0.6$ secs in our implementation) must be repeated for every value of $\rho$.

The volatility expansion introduced in Remark \ref{Kim} is easily implemented, all the terms being available in closed forms with the exception of $\Lambda(\lambda,0,T)$ which  was computed by a standard quadrature (GK) algorithm. The procedure is very fast (about $0.5\times 10^{-3}$ secs.) and since the approximation is linear in $\rho$, the estimated CVA is computed once for all values of $\rho$, as for the correlation expansion method.

\begin{table} \centering
\begin{tabular}{|r|c|c|c|c|}
  \hline
  $\rho$ & Corr. exp. & Vol. exp. & Drift adj.  & MC + control (C.I)\\  \hline
  -0.9 &    0.11780 (0.00253) & 0.11729 ( 0.00304) & 0.12215 (-0.00181) &  0.12034 (0.00009) \\
  -0.7 &    0.12712 (0.00150) & 0.12677 ( 0.00184) & 0.12970 (-0.00108) &  0.12861 (0.00010) \\
  -0.5 &    0.13643 (0.00084) & 0.13625 ( 0.00102) & 0.13769 (-0.00042) &  0.13727 (0.00012) \\
  -0.3 &    0.14575 (0.00023) & 0.14573 ( 0.00026) & 0.14615 (-0.00017) &  0.14598 (0.00013) \\
  -0.1 &    0.15506 (0.00009) & 0.15520 (-0.00004) & 0.15508 ( 0.00008) &  0.15516 (0.00014) \\
   0.1 &    0.16438 (0.00004) & 0.16468 (-0.00026) & 0.16448 (-0.00006) &  0.16443 (0.00015) \\
   0.3 &    0.17369 (0.00014) & 0.17416 (-0.00033) & 0.17437 (-0.00053) &  0.17383 (0.00015) \\
   0.5 &    0.18301 (0.00062) & 0.18364 (-0.00000) & 0.18473 (-0.00110) &  0.18364 (0.00015) \\
   0.7 &    0.19233 (0.00156) & 0.19312 ( 0.00077) & 0.19558 (-0.00169) &  0.19389 (0.00015) \\
   0.9 &    0.20164 (0.00250) & 0.20260 ( 0.00154) & 0.20692 (-0.00277) &  0.20414 (0.00014) \\
   \hline
\end{tabular}
\caption{Numerical results for varying $\rho$, $\sigma=0.1$. In parenthesis the errors with respect to the MC values and, for the MC values, the $95\%$ confidence interval length. The CIR volatility is $\eta=0.1$.} \label{Tab1}
\end{table}

\begin{table} \centering
\begin{tabular}{|r|c|c|c|c|c|}
  \hline
  $\rho$ & Corr. exp. & Vol. exp. & Drift adj. & MC + control (C.I)\\  \hline
  -0.9 &  0.04460 (0.02252) &  0.03199 ( 0.03514) &  0.07181 (-0.00468) & 0.06713 (0.00015) \\
  -0.7 &  0.06979 (0.01364) &  0.06042 ( 0.02302) &  0.08500 (-0.00156) & 0.08344 (0.00020) \\
  -0.5 &  0.09499 (0.00688) &  0.08886 ( 0.01302) &  0.10125 ( 0.00063) & 0.10188 (0.00027) \\
  -0.3 &  0.12018 (0.00247) &  0.11729 ( 0.00536) &  0.12100 ( 0.00166) & 0.12265 (0.00034) \\
  -0.1 &  0.14537 (0.00014) &  0.14573 (-0.00050) &  0.14462 ( 0.00060) & 0.14522 (0.00041) \\
   0.1 &  0.17057 (0.00025) &  0.17416 (-0.00334) &  0.17237 (-0.00155) & 0.17082 (0.00047) \\
   0.3 &  0.19576 (0.00251) &  0.20260 (-0.00432) &  0.20437 (-0.00610) & 0.19827 (0.00052) \\
   0.5 &  0.22095 (0.00689) &  0.23103 (-0.00318) &  0.24059 (-0.01275) & 0.22784 (0.00056) \\
   0.7 &  0.24614 (0.01360) &  0.25946 ( 0.00029) &  0.28087 (-0.02111) & 0.25975 (0.00057) \\
   0.9 &  0.27134 (0.02248) &  0.28790 ( 0.00592) &  0.32493 (-0.03111) & 0.29382 (0.00056) \\
   \hline
\end{tabular}
\caption{Numerical results for varying $\rho$, $\sigma=0.1$. In parenthesis the errors with respect to the MC values and, for the MC values, the $95\%$ confidence interval length. The CIR volatility is $\eta=0.3$. } \label{Tab2}
\end{table}

\begin{table} \centering
\begin{tabular}{|r|c|c|c|c|c|}
  \hline
  $\rho$ & Corr. exp. & Vol. exp. & Drift adj. & MC + control (C.I)\\  \hline
  -0.9 & 0.00005 (0.04704) & -0.05330 ( 0.10029) & 0.04566 ( 0.00132) & 0.04698 (0.00016) \\
  -0.7 & 0.03431 (0.02821) & -0.00591 ( 0.06844) & 0.05762 ( 0.00489) & 0.06252 (0.00024) \\
  -0.5 & 0.06868 (0.01433) &  0.04147 ( 0.04154) & 0.07493 ( 0.00807) & 0.08301 (0.00036) \\
  -0.3 & 0.10305 (0.00530) &  0.08886 ( 0.01950) & 0.09960 ( 0.00875) & 0.10836 (0.00049) \\
  -0.1 & 0.13742 (0.00052) &  0.13625 ( 0.00170) & 0.13361 ( 0.00433) & 0.13795 (0.00063) \\
   0.1 & 0.17179 (0.00041) &  0.18364 (-0.01143) & 0.17841 (-0.00620) & 0.17220 (0.00077) \\
   0.3 & 0.20616 (0.00528) &  0.23103 (-0.01959) & 0.23451 (-0.02306) & 0.21144 (0.00090) \\
   0.5 & 0.24053 (0.01653) &  0.27842 (-0.02135) & 0.30140 (-0.04432) & 0.25707 (0.00103) \\
   0.7 & 0.27491 (0.02987) &  0.32581 (-0.02102) & 0.37783 (-0.07304) & 0.30478 (0.00111) \\
   0.9 & 0.30928 (0.05128) &  0.37320 (-0.01263) & 0.46226 (-0.10169) & 0.36057 (0.00115) \\
   \hline
\end{tabular}
\caption{Numerical results for varying $\rho$, $\sigma=0.1$. In parenthesis the errors with respect to the MC values and, for the MC values, the $95\%$ confidence interval length. The CIR volatility is $\eta=0.5$. } \label{Tab3}
\end{table}

\begin{table} \centering
\begin{tabular}{|r|c|c|c|c|c|}
  \hline
  $\rho$ & Corr. exp. & Vol. exp. & Drift adj. & MC + control (C.I)\\  \hline
  -0.9 &  0.34222 ( 0.00937) &  0.34623 ( 0.00537) &  0.35829 (-0.00667) & 0.35160 (0.00030) \\
  -0.7 &  0.37230 ( 0.00576) &  0.37557 ( 0.00249) &  0.38192 (-0.00386) & 0.37806 (0.00036) \\
  -0.5 &  0.40238 ( 0.00292) &  0.40490 ( 0.00040) &  0.40714 (-0.00184) & 0.40530 (0.00040) \\
  -0.3 &  0.43246 ( 0.00046) &  0.43424 (-0.00132) &  0.43403 (-0.00110) & 0.43292 (0.00044) \\
  -0.1 &  0.46254 (-0.00005) &  0.46358 (-0.00109) &  0.46262 (-0.00014) & 0.46249 (0.00048) \\
   0.1 &  0.49262 ( 0.00023) &  0.49292 (-0.00006) &  0.49299 (-0.00013) & 0.49285 (0.00050) \\
   0.3 &  0.52270 ( 0.00077) &  0.52225 ( 0.00122) &  0.52517 (-0.00169) & 0.52348 (0.00051) \\
   0.5 &  0.55278 ( 0.00288) &  0.55159 ( 0.00407) &  0.55921 (-0.00354) & 0.55566 (0.00052) \\
   0.7 &  0.58286 ( 0.00605) &  0.58093 ( 0.00799) &  0.59514 (-0.00622) & 0.58892 (0.00057) \\
   0.9 &  0.61294 ( 0.00922) &  0.61027 ( 0.01190) &  0.63300 (-0.01083) & 0.62216 (0.00048) \\
   \hline
\end{tabular}
\caption{Numerical results for varying $\rho$, $\sigma=0.3$. In parenthesis the errors with respect to the MC values and, for the MC values, the $95\%$ confidence interval length. The CIR volatility is $\eta=0.1$. } \label{Tab4}
\end{table}

\begin{table} \centering
\begin{tabular}{|r|c|c|c|c|c|}
  \hline
  $\rho$ & Corr. exp. & Vol. exp. & Drift adj. & MC + control (C.I)\\  \hline
  -0.9 & 0.54936 ( 0.01904) &  0.55828 ( 0.01012) & 0.56840 (-0.01310) & 0.56840 (0.00054) \\
  -0.7 & 0.60299 ( 0.01159) &  0.61017 ( 0.00441) & 0.61459 (-0.00784) & 0.61459 (0.00064) \\
  -0.5 & 0.65663 ( 0.00519) &  0.66207 (-0.00026) & 0.66182 (-0.00453) & 0.66182 (0.00073) \\
  -0.3 & 0.71026 ( 0.00163) &  0.71397 (-0.00208) & 0.71189 (-0.00163) & 0.71189 (0.00081) \\
  -0.1 & 0.76390 ( 0.00089) &  0.76587 (-0.00108) & 0.76479 ( 0.00069) & 0.76479 (0.00087) \\
   0.1 & 0.81753 (-0.00007) &  0.81777 (-0.00030) & 0.81746 (-0.00078) & 0.81746 (0.00092) \\
   0.3 & 0.87117 ( 0.00269) &  0.86966 ( 0.00419) & 0.87386 (-0.00225) & 0.87386 (0.00096) \\
   0.5 & 0.92480 ( 0.00615) &  0.92156 ( 0.00939) & 0.93095 (-0.00689) & 0.93095 (0.00096) \\
   0.7 & 0.97844 ( 0.01078) &  0.97346 ( 0.01576) & 0.98922 (-0.01436) & 0.98922 (0.00096) \\
   0.9 & 1.03207 ( 0.01804) &  1.02536 ( 0.02475) & 1.05011 (-0.02337) & 1.05011 (0.00091) \\
   \hline
\end{tabular}
\caption{Numerical results for varying $\rho$, $\sigma=0.5$. In parenthesis the errors with respect to the MC values and, for the MC values, the $95\%$ confidence interval length. The CIR volatility is $\eta=0.1$. } \label{Tab5}
\end{table}

\begin{figure} \hspace{-1.2cm}
\includegraphics[width=14cm,height=7cm]{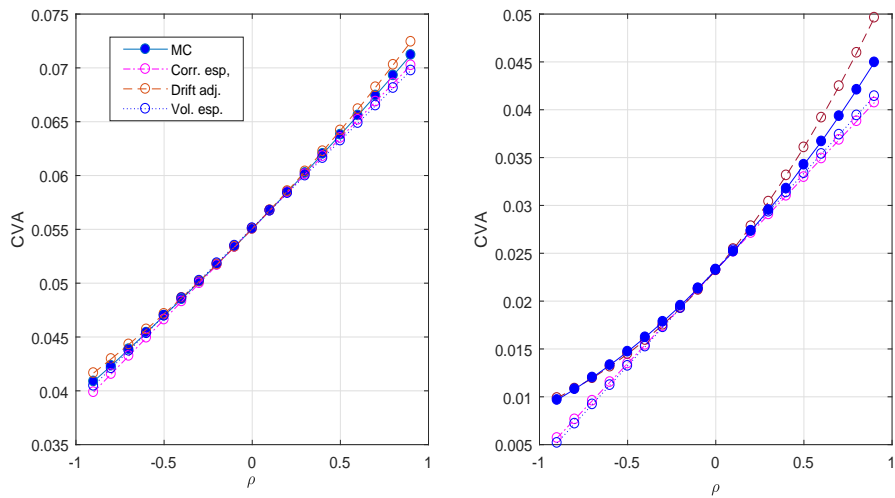}
\caption{Comparison of all methods for the set of parameters in Brigo et al. \cite{BRH18}, maturity $T=1$, parameter set 1 on the left and parameter set 3 on the right.}
\label{fig1}
\end{figure}

\begin{figure}\hspace{-1.2cm}
\includegraphics[width=14cm,height=7cm]{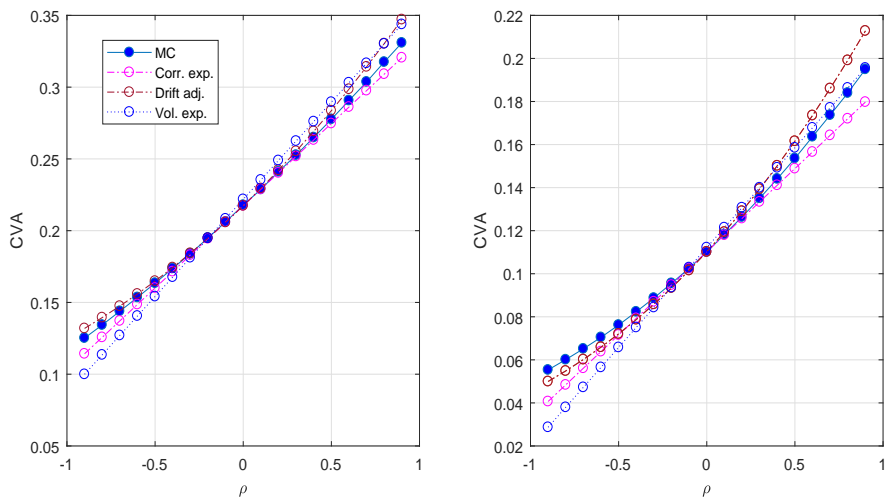}
\caption{Comparison of all methods for the set of parameters in Brigo et al. \cite{BRH18}, maturity $T=5$, parameter set 1 on the left and parameter set 3 on the right.}
\label{fig2}
\end{figure}

The approximation methods are compared to MC (with control variates) estimates, the error being defined as $\widehat{CVA}_{MC} - \widehat{CVA}_{Method}$. A positive sign indicates an underestimation of the CVA with respect to MC. In our experiments (Tables (\ref{Tab1}) to (\ref{Tab5})) we noticed that the three methods provide better approximation for small values of $|\rho|$: the correlation expansion, which is linear in $\rho$, provides a lower bound for CVA, while the drift adjustment gives a uniformly level of approximations which, however slightly, worsens as the values of $\rho$ become larger and positive (other choices of the $\lambda(t)$ tend to mitigate this effect, see \cite{BRH18}). In particular we experienced a systematic underestimation of the WWR effect for the correlation expansion method and an overestimation for the drift adjustment method, while the volatility expansion has not a definite behavior. This kind of pattern is still observed for the other parameter sets considered (see Figures (\ref{fig1}), (\ref{fig2})).

As pointed out in Remark (\ref{WWR_g1}), the contribution to the CVA due to the correlation $\rho$ is quantified by $g_1$: its behavior is reported in Table (\ref{g1_val}) and it suggests an increasing impact of WWR for the volatility of the default intensity becoming larger.

\begin{table} \centering
\begin{tabular}{cccccc}
  \hline
$\eta$  & $0.1$ & $0.2$ & $0.3$ & $0.4$ & $0.5$ \\ \hline
$|g_1|$   & $0.0466$ & $0.0898$ & $0.1260$ & $0.1532$ & $0.1719$ \\ \hline
\end{tabular}
\caption{The absolute value of $g_1$ for different volatilities $\eta$.} \label{g1_val}
\end{table}

We further compared the approximation methods on the same two sets of parameters (set 1 and 3) used in \cite{BRH18} for the CIR dynamic, see Table (\ref{paramsets}). The results for  $T=1$ and $T=5$ are reported graphically in Fig. (\ref{fig1}) and Fig. (\ref{fig2}), respectively confirming the behavior observed.

\begin{table} \centering
\begin{tabular}{rcccc}
  \hline
      &  $\lambda_0$ & $\gamma$ & $\theta$ & $\eta$  \\ \hline
Set 1 &  $0.03$      & $0.02$   & $0.161$  & $0.08$ \\
Set 3 &  $0.01$      & $0.8$    & $0.02$   & $0.2$ \\ \hline
\end{tabular}
\caption{Parameter sets.} \label{paramsets}
\end{table}

\section{Conclusions}

We considered the pricing problem for financial options subject to counterparty credit risk. The impact of a credit event is quantified by the Credit Value Adjustment, which we modeled in a stochastic intensity framework. This allows to represent the CVA as the expectation of the derivative's payoff   discounted with a rate  given by the sum of the risk-free and of the default intensity. Wrong Way Risk is accounted for by considering positive dependence between the exposure and the default event. The calculation of such a quantity may be tackled by classical Monte Carlo methods once the dynamics of the stochastic state variables  (underlying, risk-free rate and default intensity) are chosen, but it is  computationally very expensive. As an alternative to that, we proposed in this paper the correlation expansion method to evaluate CVA with WWR, when the underlying  and the intensity dynamics are respectively given by a geometrical Brownian motion and a CIR process. Finally we compared the performance of our method with that of two other semi-analytical techniques: the drift adjustment introduced in \cite{BR18} and the volatility expansion technique used in \cite{KK99}.


\begin{thebibliography}{00}
\bibitem {AL} C. Albanese,  S. Lawi, \textit{ Laplace transforms for integrals of Markov processes}, Markov Process and Related Fields, 11, 677--724 (2005).

\bibitem{AS} F. Antonelli,  S. Scarlatti, \textit{Pricing Options under stochastic volatility: a power series approach}, Finance and Stochastics, 13, 269--303 (2009).

\bibitem{ARS} F. Antonelli, A. Ramponi, S. Scarlatti, \textit{Exchange option pricing under stochastic volatility: a correlation expansion}, Review of Deriv. Research, 13, 45--73,   (2010).

\bibitem{ARS1} F. Antonelli, A. Ramponi, S. Scarlatti, \textit{Random time forward-starting options},
Int. J. of Theor. and Appl. Finance, 19, 8, (2016).

\bibitem{BCB} T. R. Bielecki, S. Crepey, D. Brigo, \textit{Counterparty Risk and Funding: A Tale of Two Puzzles.} Chapman and Hall/CRC,  (2014).

\bibitem{BR} T. R. Bielecki, M. Rutkowski, \textit{Credit Risk: Modeling, Valuation and Hedging},  Springer Finance Series (2002).

\bibitem{BJR} T. R. Bielecki, M. Jeanblanc, M. Rutkowski, \textit{Valuation and Hedging of Credit Derivatives}, Lecture notes CIMPA- UNESCO Morocco School,  (2009).

\bibitem{BR18} D. Brigo, F.Vrins, \textit{Disentangling wrong-way risk: pricing credit valuation adjustment via change of measures}, European Journal of Operational Research, 269, 1154--1164, (2018).

\bibitem{BMP13}D.Brigo, M.Morini, A.Pallavicini, \textit{ Counterparty Credit Risk, Collateral and Funding: With Pricing Cases For All Asset Classes},Wiley, (2013).

\bibitem{BRH18} D. Brigo ,T. Hvolby, F.Vrins,\textit{Wrong-Way Risk adjusted exposure: Analytical Approximations for Options in Default Intensity Models}, to appear in  WSPC Proceedings (2018).

\bibitem{CD03} E. Canabarro, D. Duffie,\textit{Measuring and marking counterparty risk?, Asset/Liability Management of Financial Institutions},Euromoney books,(2003).

\bibitem{CPV14} A. Capponi, S. Pagliarani, T. Vargiolu, \textit{Pricing vulnerable claims in a Levy driven model}
Finance and Stochastics, 18, 755 --789, (2014).

\bibitem{C13} U. Cherubini, \textit{Credit valuation adjustment and wrong way risk}, Quantitative Finance Letters, 1, 9--15, (2013).

\bibitem{CL03} U. Cherubini, E. Luciano, \textit{Pricing Vulnerable Options with Copulas}, Journal of Risk Finance, vol. 5, 27--39, (2003).


\bibitem{DS99} D. Duffie, K.J. Singleton,\textit{ Modeling term structures of defaultable bonds}, Review Financial Studies, 12 , 687--720,(1999).

\bibitem{Fard15} F.A. Fard, \textit{Analytical pricing of vulnerable options under a generalized
jump-diffusion model}, Insurance Mathematics and Economics, 60, 19--28, (2015).

\bibitem{G14} P. V. Gapeev, \textit{Some extensions of Norros' lemma in models with several defaults. Inspired by Finance}, The Musiela Festschrift. Kabanov Yu. M., Rutkowski M., Zariphopoulou Th. eds. Springer, 273--281, (2014).

\bibitem{GJLR10} P. V. Gapeev, M. Jeanblanc, L. Li, M.Rutkowski, \textit{Constructing random measures with given survival processes and applications to valuation of credit derivatives}, Contemporary Quantitative Finance, Essays in Honour of Eckhard Platen. Chiarella, C., Novikov, A. eds. Springer, 255--280, (2010).

\bibitem{GY16} P. Glasserman, L. Yang \textit{Bounding wrong way risk in cva calculations},Mathematical Finance, 28, 268--305,(2016).

\bibitem{Gr12} J.Gregory, \textit{Counterparty credit risk and credit value adjustment}, Wiley, (2012).

\bibitem{HW95} J. Hull, A. White, \textit{The impact of default risk on the prices of options and other derivative securities}, Journal of Banking \& Finance, 19, 299--322,(1995).

\bibitem{HW12} J. Hull, A. White, \textit{CVA and Wrong Way Risk},Financial Analyst Journal, 68, 58--69,(2012).

\bibitem{Kao16} L.J. Kao,\textit{Credit valuation adjustment of cap and floor with counterparty risk: a structural pricing model for vulnerable European options}, Review of Deriv. Research, 19, 41--64,(2016).

\bibitem{JT95} R. Jarrow, S. Turnbull, \textit{Pricing derivatives on financial securities subject to credit risk}, Journal of Finance, 50, 53--85, (1995).

\bibitem{JS87} H. Johnson, R. Stulz, \textit{The Pricing of Options with Default Risk}, Journal of Finance, 42, 267-280,(1987).

\bibitem{KA03} P. Knaber, L. Angermann,\textit{Numerical Methods for Elliptic and Parabolic Partial Differential Equations}, Springer,(2003).

\bibitem{Kl96} P. Klein,\textit{Pricing Black-Scholes options with correlated credit risk},Journal of Banking \& Finance, 20,1211--1229,(1996).

\bibitem{KIn99} P. Klein, M. Inglis,\textit{Valuation of European options subject to financial distress and interest rate risk}, Journal of derivatives, 6, 44--56, (1999).

\bibitem{KIn01} P. Klein, M. Inglis,\textit{Pricing vulnerable European option's when the option payoff  can increase the risk of financial distress},Journal of Banking \& Finance, 25, 993--1012,(2001).

\bibitem{KK99} Y.J. Kim, N. unimoto ,\textit{Pricing Options under Stochastic Interest Rates: A New Approach.} Asia-Pacific Financial Markets, 6, 49--70, (1999).

\bibitem{HW2016} H. Niu, D. Wang,  \textit{Pricing vulnerable options with correlated jump-diffusion processes depending on various states of the economy}, Quantitative Finance,16, 7, 1129--1145, (2016).

\bibitem{Lando98} D.Lando, \textit{On Cox Processes and Credit Risky Securities}, Review of Derivatives Research, 2, 99--120 (1998).

\bibitem{LLPB} T. Lepage, S. Lawi, P. Tupper, D. Bryant, \textit{Continuous and tractable models for the variation of evolutionary rates}, Mathematical Biosciences, 199, 216--233 (2006).

\bibitem{N} A. Nikeghbali, \textit{An essay on the general theory of stochastic processes}, Probability Surveys, 3, 345--412, (2006).

\bibitem{PR10} M. Pykhtin , D. Rosen, D., \textit{Pricing Counterparty Risk at the Trade level and CVA Allocations}, Journal of Credit Risk, 6, 3--38 (2010).

\bibitem{PP} A. Prayoga, N.Privault, \textit{Pricing CIR Yield Options by Conditional Moment Matching}, Asia-Pacific Financial Markets, 24, 1 , 19--38 (2017).

\bibitem{Rab89} R. Rabinovitch, \textit{Pricing Stock and Bond Options when the Default-Free Rate is Stochastic}, The Journal of Financial and Quantitative Analysis, 24, No. 4, 447--457,  (1989).

\bibitem{RS12} D.Rosen,D.Saunders,\textit{CVA the wrong way}, Journal Risk Management Financial Institutions, 5, 252--272,(2012).

\bibitem{TW14} L.Tian, G.Wang, X.Wang, Y.Wang,\textit{Pricing vulnerable options with correlated credit risk under jump-diffusion processes}, Journal of Futures Markets,
34, 957--979,(2014).

\bibitem{ZP07} S.Zhu, M.Pykhtin,\textit{A Guide to Modeling Counterparty Credit Risk}, GARP Risk Review, July/August (2007)

\end{thebibliography}
\end{document}